\documentclass[final,5p,times,twocolumn,authoryear]{elsarticle}

\usepackage{amssymb}
\usepackage{lipsum}
\usepackage{graphicx}
\usepackage{dcolumn}
\usepackage{bbold}
\usepackage[version=4]{mhchem}
\usepackage{amsmath}
\usepackage{physics}
\usepackage{bm}
\usepackage{color}

\begin{document}

\begin{frontmatter}

\title{On the  Electric Dipole Moment of the Neutron and its  Quantum Uncertainty}

\author[first]{Octavio Guerrero}
\affiliation[first]{organization={Department of Physics, University of Texas at Austin},
            city={Austin},
            postcode={78712}, 
            state={TX},
            country={USA}}

\author[second]{Libertad Barrón-Palos}
\affiliation[second]{organization={Instituto de Física, Universidad Nacional Autónoma de México},
            addressline={POB 20-364}, 
            postcode={01000}, 
            state={CDMX},
            country={Mexico}}

\author[third]{Daniel Sudarsky}
\affiliation[third]{organization={Instituto de Ciencias Nucleares, Universidad Nacional Autónoma de México},
            addressline={POB 70-543}, 
            postcode={04510}, 
            state={CDMX},
            country={Mexico}}

\begin{abstract}
The continued interest in placing bounds on the neutron's  Electric Dipole Moment (EDM) is due to the implications regarding the characteristics of the strong interaction and, in particular, its behavior under the CP  symmetry. In this work, we discuss the apparent tension resulting from the discrepancy of about 13 orders of magnitude between the current bounds and the expected quantum uncertainty in the relevant quantity. We offer a resolution of the  ``puzzle"   in terms of the notion of a weak measurement, using a version of the corresponding formalism adapted to consideration of the nEDM experiment at the Spallation Neutron Source at the Oak Ridge National Laboratory.  
\end{abstract}


\end{frontmatter}

\section{Introduction}
\label{introduction}

The search for an indication of   a non-zero value of the neutron EDM\footnote{At a much higher  level than that  which could  be accounted for  as a result of  the CP  violation in the   electroweak interactions.} 
is motivated by the fact that  
it would represent a new source of CP symmetry violation with origins in nontrivial topological features of the QCD vacuum \citep{QCD_Vaccum}. That, in turn, has implications on theories beyond the standard model of particle physics, as in the standard model itself \citep{ENGEL201321, Hooft}. 

In the quest to study this quantity,  quite significant projects have been developed relying on multiple techniques that have achieved 
remarkable levels of control of the statistical and systematical errors \citep{Abel, ExperSNSImp}. Recently, using the Paul Scherrer Institut's ultracold neutron source, the nEDM collaboration has reported the lowest bound on neutron EDM value to date $d_n= (0.0\pm1.1_{stat}\pm0.2_{sys})\times 10^{-26} \;e\cdot$cm \citep{Abel2020}.  Moreover, the nEDM experiment at the Spallation Neutron Source of the Oak Ridge National Laboratory (nEDM@SNS) seeks to study the neutron EDM  up to an accuracy of $\mathcal{O}(10^{-28} \;e\cdot$cm) when the experiment starts working at maximum capacity \citep{Leung_2019}.

\section{Statement of the problem}
In this paper, we will discuss a tension that is already present between the uncertainty ( or  {\it``dispersion"} \citep{sakurai}) associated with the neutron EDM and the experimentally measured results.
The analysis starts by considering the order of magnitude of the neutron mean square radius, which can be extracted from simple back-of-the-envelope calculations but can also be inferred from the deep inelastic scattering of electrons on protons and the well-known facts about the extreme similarity between protons and neutrons (as represented, for instance, in the SU(2)—Isopspin symmetry of the strong interactions).
 
We should start by clarifying that the question we want to analyze in this manuscript refers to the compatibility of the basic aspects of quantum theory and the actual practice, including the results reportedly obtained by our experimental colleagues. In other words,  we are not so much concerned with the creativity and resourcefulness displayed by the physicist who carries out certain extremely high precision determinations but the extent to which our understanding of such experiments fits into the general scope of physical systems'  characterization and the corresponding limitations implied by quantum theory\footnote{ There are of course many other experiments besides the one we are focused on here where remarkable and often baffling levels of precision and sensitivities have been achieved, (for instance the sensitivities to surprisingly small changes in the  ``distance between mirrors"    arrived at in  LIGO,  or the precision measurements on the anomalous magnetic moment of the electron, to name a few),  and our focus on the problem at hand is not meant to demerit those achievements by comparison.  However, many of those experiments do not seem to imply, even at first sight, the kind of puzzle we face here, and some which might (we have not made an exhaustive analysis of all experiments that might fall in this category) can perhaps be treated in an analogous way as the present case. } 
 
 Now, let us turn back to the problem at hand. The effective radius of the neutron is of the order of femtometers \citep{NeutronRadii} $\mathcal{O}(10^{-13}$ cm), 
 thus, by considering a very simplified characterization of the neutron in terms of  its valence    quarks (two   down  quarks   with charge    $-1/3  \;e $ and   an up quark with charge $  2/3  \;e $ ) and  the  corresponding  explicit  form of the electric  dipole moment operator: 

\begin{equation}
    \hat{\vec{D}}=\sum_i^3 q_i \hat{\vec{x}}_i,
    \label{nEDM Definition}
\end{equation}

where $q_i$ are the respective charges of the  quarks 
and $\hat{\Vec{x}}_i$ the positions of each one of them (for the sake of simplicity and unless clarity requires it, we will drop the operator hats notation).

We can  easily  make a back-of-the-envelope estimation of   the value of  the  nEDM uncertainty, finding  it  to be  about 
$ \Delta  |D| \sim  \mathcal{O}(10^{-13} \;e\cdot$cm).  
That represents a discrepancy of  13 orders of magnitude between the precision of the last reported value of this quantity and the intrinsic quantum uncertainty in the same quantity. This seems quite puzzling when one considers that, as per the usual expectations from quantum theory, the quantum uncertainty  (in the corresponding system's quantum state) represents, among other things,  a fundamental limit to the precision with which any observable can be measured for a system in the corresponding state.  
 
So, what are we to make of the situation at hand? Let us consider the most straightforward replies one might offer.

  i)  One might argue   that such analysis  is too  simplistic and contemplate,   for instance,  a simple  harmonic  oscillator  with  frequency $ \omega $  and mass  $ M$, which is prepared in its ground  state, in   which the position $X$  
has an expectation value  $ \langle  \hat X \rangle  =0 $  and the quantum  uncertainty  is  $ \Delta X =\hbar/ (2 M\omega)^{1/2}$. One might argue that it is true that  $\Delta X$ represents the degree to {\bf which the particle's} position is {\bf ill-defined}  in the initial state  $ | 0 \rangle$;  however, that does not mean we cannot perform an experiment     
measuring the position to a higher accuracy. In fact, we can certainly do that and, in principle, measure the position of the system to an arbitrary accuracy $ \delta X$  (at least in the context of non-relativistic quantum  mechanics\footnote{Inclusion of relativistic considerations indicate that we should not be able to measure the position of a particle with a precision that exceeds the particle Compton's wavelength, but that let us ignore this issue for the moment.}).  All that quantum mechanics tells us is that 
our prediction of what the value resulting from the measurement of the position must be taken as uncertain to a level  $ \Delta X $,   indicating that if we repeat the experiment a large number of times (with identically prepared systems), we will obtain a series of results whose mean value is $ \bar X = 0$ and with a statistical dispersion given  
by   $ \Delta X $.  However, in each one of the measurements, the position might end up being well determined within an uncertainty  $ \delta X$  ( corresponding to the accuracy of the measuring device), and thus, there is no conflict at all between quantum theory and the fact that we have measured   $X$ with an accuracy that far exceeds $ \Delta X$.
Note, however, that, as a result of such measurements, the quantum state of each oscillator would have changed to one ``centered about some definite value of the position, say $X_i$" with an uncertainty $ \delta X << \Delta X $  and the collection of values obtained in the ensemble of measurements  $\lbrace X_1, X_2,...... X_N\rbrace$ would  display a  statistical dispersion  
$ \Delta X $. In particular, all the harmonic oscillators would now be in states quite different from the ground state,  and thus, their corresponding energy expectation values would be higher than $  (1/2) \hbar \omega$.
  
 In the situation at hand, it is quite clear that what was described above cannot be what is going on. There are various reasons for that. First note that if the experiment involves a large number of neutrons corresponding to a repetition of the measurement of nEDM  (with accuracies of order   $\mathcal{O}(10^{-28}\;e\cdot$cm$)$), we would just as with the example of the harmonic oscillator,  obtain an ensemble of different results with a statistical dispersion of order $ \Delta  |D| \sim  \mathcal{O}(10^{-13}\; e\cdot$ cm$)$ (and our experimental colleagues would not be able to report the result they do).   Furthermore, as we just noted,    in the case of the harmonic oscillators,  such kind of measurement (with accuracies that are much higher than the quantum uncertainty of the original state)  led to changes in the state of the system, and,  when the original system was in the ground state,  this implied an increase in the expectation value of the system's energy. The energy scales controlling the internal structure of a neutron are of the order of MeV's, which is an enormous scale compared with the energies that the experimental devices used in the type of experiments under consideration might ``transfer''  to the neutrons. In fact, neutrons are the ground state of that type of quark arrangement,  and the higher excitation levels correspond to the particles such as $  \Delta^{0}, $   as well those known as resonances $N^{*}$, etc.  All these particles mentioned have significantly higher masses compared to a neutron. It's exceptionally challenging even to contemplate the possibility that the internal structure of a neutron might undergo substantial modification due to the relatively weak electric and magnetic fields utilized in these experiments. Consequently, drawing an analogy with the type of measurement we considered in the context of the harmonic oscillator breaks down entirely. In the case of the harmonic oscillator, the measurement profoundly alters the system's state, whereas in this scenario, such a drastic transformation cannot occur.

 The considerations above also serve to show the difficulties one would face if one were to attempt to measure the expectation value of nEDM by brute force, namely by carrying out  (ordinary)  strong measurements of the hermitian operator  $  \hat{\vec{D}}$  for a large number of neutrons then arranging the results, say, on a suitable histogram and computing the corresponding average value. At first sight, one might see no difficulty in obtaining the desired expectation value to whatever accuracy one desires, as, of course, there is,   in principle, no limitation provided by quantum theory to the precision with which one might measure a  single observable.   Note that on the one hand, this would involve the formidable task of measuring each neutron's  EDM with a precision of no less than  $\mathcal{O}(10^{-26})\;e\cdot$cm,   for otherwise, we would not be able to reach the desired accuracy, and given that the quantum uncertainty is about  $10^{12}$ times larger, our histogram would involve billions of slots.  On the other hand, in attempting to proceed in this manner, we face an even more serious and fundamental problem.,  namely that those measurements would modify the internal structure of the neutron (which, as we noted, is something that requires rather large energies),   we would end, after each measurement, with something other than a neutron,  i.e., something like   $ \Delta^{0}, $  or $N^{*}$, and more often a multi-particle state involving not only nucleons but other kinds of hadrons like pions,  kaons,  etc. as well as photons. That is, we would not have only neutrons anymore and would then be making a serious mistake if we claimed that what we had measured was the nEDM or even its expected value. We thank an anonymous referee for asking a question that led us to note this point.  In other words, quantum theory by itself indeed places no limitation on the precision with which one might measure a  single observable,  but at the same time, the theory indicates that in doing so, we would, in general, change the state quantum of the system in possibly drastic ways. In our case, the expected change would involve transforming the neutron into something else.  Thus, this approach cannot provide a satisfactory resolution of the puzzle.

ii)  Could it be that we are instead simply overestimating the nEDM uncertainty? Could it truly vanish? For the latter to be the case,  the neutron should be an eigenstate of the electric dipole moment operator, and that seems rather problematic\footnote{We might dismiss the possibility given that at the electro-weak level, the CP symmetry is violated, and this should induce a non-vanishing expectation value for the nEDM, however as that effect is known to be minuscule, even compared with the tight bounds we are considering here, we will from now on simply ignore the electro-weak CP violation in the discussion.}, as there seems to be no reason whatsoever that could account for that.

We will now consider two kinds of analysis that offer strong evidence against that possibility and support our original order of magnitude estimates.
The first involves considering the correlations that must be present in the wave function characterizing the constitutive parts of the neutron. Such  correlation  evidenced  by   the condition imposed by the CP symmetry on the expected value of the neutron EDM is:

\begin{equation}
    \expval{{\vec{D}}}=q_1\expval{{\vec{x}}_1}+q_2\expval{{\vec{x}}_2}+q_3\expval{{\vec{x}}_3}=0.
    \label{nEDM CP Condition}
\end{equation}

where, for simplicity, we have considered just the valence quarks and treated the two  $u$    quarks as non-identical particles.

Furthermore, these correlations cannot be trivial, i.e., even though the sum of the charges of the quarks that compose the neutron equals zero $\sum_i^3q_i=0,$ the expectation values of the positions of the three quarks cannot be identical. Otherwise, the scattering experiments that have measured the mean squared radius of the proton (considering the strong resemblance between both nucleons)   would have resulted in a much smaller absolute value of this quantity  \citep{Abrahamyan_2012,Kurasawa}. 
 Therefore, the quantum state of the system must entangle the positions of the quarks to ensure the condition \eqref{nEDM CP Condition}. 

Note that  we work under the assumption of the exact validity of the CP symmetry  (so $\expval{{\vec{D}}}=0 $ ), the quantum uncertainty of  the neutron EDM is:   

\begin{equation}
    \Delta  |D|=\sqrt{\expval{{\vec{D}}^2}},
    \label{nEDM Uncertainty}
\end{equation}

Using the expression \eqref{nEDM Definition}, the right-hand side of the last equation can be expressed as 

\begin{eqnarray}
\expval{{\vec{D}}^2}&&=(q_1)^2\expval{{\vec{x}}_1^2}+(q_2)^2\expval{{\vec{x}}_2^2}+(q_3)^2\expval{{\vec{x}}_3^2}+2q_1q_2\nonumber
\\&&\times\expval{{\vec{x}}_1\cdot{\vec{x}}_2}+2q_1q_3\expval{{\vec{x}}_1\cdot{\vec{x}}_3}+2q_3q_2\expval{{\vec{x}}_3\cdot{\vec{x}}_2}.\nonumber
\\&&
\label{Right-hand Side Decomposed}
\end{eqnarray}

Consider now the previous expression's fourth, fifth, and sixth terms. They have the form $ 2q_iq_j\langle \vec{x_i}\cdot\vec{x_j}\rangle$.
At this point, we start by noting the  
mathematical inequality  $\langle \vec{x_i}\cdot\vec{x_j}\rangle  \leq  \sqrt{ \langle ||\vec{x_i} ||^2 ||\rangle  \langle ||\vec{x_j} ||^2 \rangle }$ where the equality would only be achieved if there is a complete correlation between the quantities involved. 
It's challenging to imagine complete correlations because they would only apply to the full wave function, which in our simplified model involves the three valence quarks. Thus, when the position of the third quark is integrated over, the density matrix for the remaining two particles would naturally encode a strong correlation. However, in general, this correlation ought to undergo a certain level of degradation as compared to the very rigid correlations in the state of the complete system \footnote{This feature can be illustrated even at the classical level by considering a set of billiard tables set initially with the balls in identical positions,  having a player hit the white ball in each table imparting in all cases the same fixed energy to that ball with the only quantity that differs from table to table being the initial direction of the hit.  Under these conditions, there will be a complete correlation among the conditions of the three balls in all tables,  with the entire collection described in terms of a one-parameter family (the initial angle of the billiard stick). However, if we decide to limit the consideration to only two of the balls (ignoring the white ball in all cases, for instance) at any time, the correlation between the states of the other two balls will be less than perfect,  i.e., it would have been degraded by ignoring relevant degrees of freedom.}.  
  
A more physical argument might be brought to bear in the discussion, which is based on the property of asymptotic freedom of   QCD. That indicates that as the separation between the quarks decreases and thus the relevant energy scales of the QCD interaction increase,  the strength of the interaction decreases, and thus the force responsible for the neutron's internal structure and thus, the correlations present in the wave function of its constituents can be expected to decreases,  and the quark-quark relative position correlation must become smaller than at large distances. The transition scale might, therefore, be estimated to correspond to energies of the order of, say, $10\;$GeV,  corresponding to an inter-quark separation of order  $ 10^{-14} \;$cm. This leads us to estimate\footnote{We thank  Alejandro Perez for this observation.}  $\langle \vec{x_i}\cdot\vec{x_j}\rangle \sim10^{-14 } cm^2 \sim 10^{-1}  \sqrt{ \langle ||\vec{x_i} ||^2 \rangle  \langle ||\vec{x_j} ||^2 \rangle }$ .

As the valence quarks can be taken as simply confined to a region of the size of the neutron radius, the terms of the form $ \expval{\vec{x_i}^2}$ have to be about the order of the neutron radius. All this  supports  our  original estimate of
$\Delta  |D_n|$ that can not be much smaller than that implied by the mean square radius of the neutron $(1/9\,e + 1/9\,e  + 4/9\,e)\times 10^{-13}\;$cm $\sim \mathcal{O}(10^{-13} \;e\cdot$ cm$)$. In this sense, there is a manifest discrepancy of $12-
13$ orders of magnitude with respect to the measured values at the Paul Scherrer Institut's experiment and $15$ orders of magnitude for the nEDM@SNS collaboration expected results. 

At this stage, we will accept that order of magnitude estimate and then try to confront the seemingly paradoxical situation described at the start of this section. 

\section{A proposal for resolving the  puzzle}
Once we face this huge discrepancy between what, in principle, should be possible to measure and what the experimental groups have measured, it is clear the need to try and clarify what is going on.  We are either confronting some sort of misunderstanding or a serious  
problem with the theory at its most basic level. Here, we will argue that it is the former and that, in fact, what is needed is 
the recognition that the relevant type of experiments are embodiments of what is called ``a weak measurement"  and that by doing so,  the mystery is completely resolved.     In  short,  a  weak measurement  is  a  type of experiment  in which  one  focuses  on   performing an ordinary measurement of  a  certain quantity  (referred to  as the  ancilla  or auxiliary observable),  which  is  only indirectly  related to  the quantity of interest  $O$  and
which yields  direct information  about the expectation  value of  that quantity  in the original   state of the  system  
$|\Psi \rangle $, namely 
$\langle  \Psi |\hat    O | \Psi  \rangle$,  without hardly disrupting the initial state of the system of interest $|\Psi \rangle $.   The notion was introduced in \citep{Aharonov} and further developed in several following works. For a recent review \citep{Duck}.

We will illustrate the idea as applied to the situation at hand by focusing on the experiment of the nEDM@SNS collaboration\footnote{A model for the Paul Scherrer Institut collaboration experiment can also be considered.} making use of the weak measurement formalism. The objective is not to offer a precise characterization of the experiment but to present an idealized version showing how the main ideas work in a toy model situation, which is, however,   sufficiently close to that of the actual experiments in question. That means we will take the liberty to modify the setups involved in several ways, and we will use several simplifications in the treatment that, although not strictly rigorous, will allow for a rather complete description of the relevant issues.

 A  more precise analysis is, in principle, possible, although quite likely impractical,   due to the sheer complexity of the actual experiments and the concomitant theoretical characterization that such an endeavor would entail. However, it is worthwhile describing in broad terms how we envision such realistic analysis could be carried out. In order to do so, we next offer a brief but broadly accurate description of the situation we are concerned with. 

\subsection{A  more  detailed description of   the experimental setup}

The experiment \citep{Leung_2019, GOLUB19941} is based on the fact that if there was a non-zero nEDM ($\vec{D}\neq 0$),  the precession frequency of the neutron spin would be affected when an electric field was applied. More specifically, the associated spin precession when the particle is immersed in constant electric and magnetic fields would be given by the  Larmor frequency. When  $ \vec E$ and $ \vec B $ are parallel, the Larmor frequency is 

\begin{equation}
    \hbar \omega_{\uparrow\uparrow}=2\abs{\mu_n B + D E}, 
    \label{eq:parallel}
\end{equation}

where  $E = || \vec E ||$ and  $B = || \vec B ||$, $\mu_n$ is the magnetic dipole moment of the neutron. On the other hand,  the corresponding frequency  when the electric field is reversed (antiparallel fields) is

\begin{equation}
    \hbar \omega_{\uparrow\downarrow}=2\abs{\mu_n B - D E}, 
    \label{eq:antiparallel}
\end{equation}

The Comparison of these quantities in the two cases would serve to determine   $\vec{D}$. However, even though the measurement principle is simple, the experiment is not simple at all. For an nEDM of the order of the current best limit, $\vec{D}\sim 10^{-26}\;e\cdot$cm \citep{Abel2020}, the Larmor frequencies in equations \eqref{eq:parallel} and \eqref{eq:antiparallel} differ by only $10^{-7}\;$Hz, considering a typical experimental electric field of 10 kV/cm. This difference is nothing more than the shift in Larmor frequency that a neutron precessing in a constant magnetic field of 1 $mT$ (or $10 G$) would experience if the field has fluctuations of the order of a few $fT$. Measurement of magnetic fields to that level can only be achieved via a co-magnetometer, which is a nuclear or atomic species with well-known magnetic dipole moment and electric dipole moment that can be considered zero for practical purposes\footnote{This can happen in atoms like ${}^{3}$He due to the atomic Schiff screening effect \citep{Schiff}.}. If the co-magnetometer species is set to precess together with neutrons in a homogeneous magnetic field, the changes in its Larmor precession frequency, which can be accurately determined using a SQUID \citep{Gallop}, can be used to monitor changes in the holding magnetic field. The nEDM@SNS experiment will use ${}^3$He as a co-magnetometer, which, in addition to providing a sensitive probe for magnetic field fluctuations, can also provide a means to measure relative neutron precession frequency. The capture of neutrons on ${}^3$He at low energies proceeds through the ground state of ${}^4$He, which has nuclear spin $J=0$. This produces a strong spin dependence in the $n+{}^3$He$\rightarrow p+{}^3$H nuclear reaction; in fact its cross section is given by

\begin{equation}
    \sigma_{n-{}^3\textnormal{He}}(v)=\frac{\sigma_0 v_0}{v}(1+P_{{}^3\textnormal{He}})
    \label{3He-sigma}
\end{equation}

\noindent where $\sigma_0=5333\;$b is the capture cross section at the thermal neutron velocity $v_0=2200\;$ m/s \citep{Sears} and $P_{{}^3\textnormal{He}}$ is the polarization of the spins of the ${}^3$He atoms with respect to the direction of the neutron spins. The precession frequency of neutrons or ${}^3$He in a magnetic field $B$ is $\omega_i=\gamma_i B$, with $\gamma_i$ the gyromagnetic ratio of the corresponding species\footnote{In fact neutrons and ${}^3$He have gyromagnetic ratios that differ by only about 10\%.}. Each species precesses at its own rate in the same magnetic field, with their spins oscillating between parallel and antiparallel, and the capture cross section for the nuclear reaction among them also oscillates according to equation \ref{3He-sigma}. The rate of occurrence of the nuclear reaction, and therefore the relative angle between neutron and ${}^3$He spins, can be measured in the experiment since the produced charged particles, $p$ and ${}^3$H, produce detectable light on the superfluid ${}^4$He bath in which both, $n$ and ${}^3$He, are immersed. Thus ${}^3$He as a co-magnetometer, provides information on very small fluctuations in the magnetic field while also providing a relative neutron precession frequency, which can be compared in the two configurations of an additional electric field $\vec{E}$, parallel or antiparallel to the magnetic field $\vec{B}$. Here, we should note that the degree of alignment between the spins of the two species is monitored by observing the photons generated in connection with the nuclear reaction, which, in turn, proceeds when the corresponding spins are antiparallel.  Thus, one can naturally consider that the actual measurement, namely the place where the macro-objectification takes place is in the (macroscopic) photon detectors interacting with the electromagnetic field in the relevant region of spacetime.  

The measurement mode described above is not the only one on which the nEDM@SNS experiment will rely. To achieve the most stringent limits, the so-called ``spin dressed'' method will be used \citep{Eckel}. This method is based on the fact that the spin precession frequency (either neutron or ${}^3$He) can be altered using a time-varying magnetic field perpendicular to the direction of the original homogeneous magnetic field. The precession frequency is affected by the difference in the effective gyromagnetic ratio of the species, $\gamma_i'$, which scales from the standard ratio $\gamma_i$ by a factor that depends on the frequency and amplitude of the applied time-varying field ($B_{RF}$ and $\omega_{RF}$). The scaling or ``dressing'' of the spin is characterized by a zeroth-order Bessel function, $\gamma_i'=\gamma_i J_0(\gamma B_{RF}/\omega_{RF})$. It is possible to find critical points where the Bessel functions of both species ($n$ and ${}^3$He) have the same value and thus do their precession frequencies, so by measuring relative frequencies (through scintillating light produced by reaction products) in the vicinity of critical points, in the two electric field configurations, it is possible to extract the value of the nEDM.

 A detailed experiment description can be found in \citep{Leung_2019}.
 
  \subsection{A  schematic analysis of the  actual  experiment as  a weak  measurement}
  
As we have seen, the experiment involves a large number of subsystems,  all of which should, in principle, be given a quantum treatment,  while in order to avoid having to confront the measurement problem in quantum theory \citep{Measurement-Problem}, we prescribe a reasonable place to set the ``Heisenberg cut"  and stipulate that certain suitable elements act as measuring devices,  which generate  ``effective classical outputs''.   In the experimental setup described above, it seems quite natural to single out the light detectors   (i.e.,  the photo-multipliers)  as the actual measuring devices that can be considered as constantly monitoring the quantum state of the electromagnetic  (EM) field in the corresponding frequency range. Moreover, we could take the externally applied electric and magnetic fields and give them a   
classical treatment (or alternatively take them to be described quantum mechanically by a suitable coherent state of the extremely low-frequency range modes of the   EM field).  Then we would consider a quantum system made up corresponding to the second quantized neutron field,  an effective second quantized  ${}^3$He ,  $p$, and ${}^3$H  fields, as well as the EM  field modes in a suitable energy range interacting (via a suitable effective Hamiltonian describing the nuclear reaction and the photon emission)   in the presence of the  (very low energy) external  EM fields. The initial state would correspond to a single neutron in a well-localized wave packet and a spin aligned on the appropriate direction, a single ${}^3 $He nucleus, the vacuum state for the   $p$,  $ {}^3$H fields,  the vacuum for the  EM high energy modes,  and unexcited state for the detectors. 
One would then write the initial state of the whole system, including the detectors  (described as low-level Unruh de-Witt detectors,   interacting with the EM field),   and consider the evolution of such state for a certain period of time and then compute the amplitude and then the probability for one of the detectors to be excited (following a  Von-Newman measurement scheme when one assumes that detectors are at the end of the experiment (i.e., at the appropriate time)  either excited or unexcited). That result would depend on various quantities,  including, as we will see in the next section, the expectation value of the EDM of the neutron. As is the case with any  weak  measurement, a single experiment will provide  very little information  about the  quantity  of interest, 
but a large sequence of identical experiments can provide arbitrarily accurate information about it.

As is well known,  one can move the  ``Heisenberg cut"  in a large number of ways, obtaining,  in practice, equivalent results.  In that fashion, one could simplify or complicate the analysis.  One should, however, avoid moving the  ``cut"  to the point that one would end describing a strong measurement of the neutron  EDM, as that would misrepresent what is really taking place experimentally. The point we have been making is that the experiments in question could not possibly correspond to such a strong measurement. 

\subsection{A  simplified  analysis of the experiment as a  weak measurement}
  
As a first simplification, in our analysis, we will focus only on the neutron, its internal degrees of freedom, and the spatial orientation of its spin.   We will treat the external electromagnetic field classically and assume that the magnetic field $\vec{B}=0$ so that the precession of the spin direction only depends on $\vec{D}$ and $\vec{E}$. Then, if $\vec{D}=0$, there will be no precession. 
The next consideration we will take is that the direction of polarization of the neutrons will be fixed in the same plane as that of the $\ce{^{3}He}$ atoms polarization, but with an angular difference of zero so that they are fully parallel.  Thus, for the capture reaction $n + \ce{{}^3He}\rightarrow p+ \ce{{}^3H}$ to occur with a probability different from zero, the neutron would need to acquire a component of its spin in the opposite direction.

To carry out a simplified version of the weak measurement of the neutron's EDM (Electric Dipole Moment), we will distinguish between the neutron's internal degrees of freedom that determine the magnitude of the EDM and the degrees of freedom that characterize the spatial orientation of the neutron.

This approach relies on a simplified form of the Wigner-Eckart theorem, known as the \textit{Projection Theorem} \citep{sakurai}. According to this theorem, for any vector operator $\hat{\Vec{V}}$ and $j\neq0$,

\begin{equation}
\bra{\alpha',j m'}V_q\ket{\alpha,j m}=\frac{\bra{\alpha',jm}\hat{\Vec{J}}\cdot\hat{\Vec{V}}\ket{\alpha,jm}}{j(j+1)}\bra{jm'}J_q\ket{jm},
\end{equation}

where $\hat{\Vec{V}}$ and $\hat{\Vec{J}}$ are in the spherical basis, and $q=(-1,0,1)$. Since we are focusing on just one of these components, we can select it as our zeroth entry.

In order to proceed with our analysis, it is convenient to introduce the operator defined by $\hat{\cal{D}}$=$(\hat{\vec{S}} \cdot \hat{\vec{D}}+\hat{\vec{D}} \cdot \hat{\vec{S}})/2$. We note that it is a Hermitian operator that codifies the magnitude of the electric dipole moment while ignoring its orientation (relying on the  Wigner-Eckart theorem). 

In fact, by leveraging the vectorial nature of the electric dipole moment operator $\hat{\vec{D}}$ (for clarity, we have reintroduced the hat notation over the operators in the following calculations) and the previous theorem, we can express the expectation value of each of its components in the following manner:

\begin{equation}
\expval{\hat{\vec {D_i}}} =  \frac{\expval{\hat {\cal D}}}{j(j+1)}\expval{\hat {\vec {J_i}}}
\label{effective-id}
\end{equation}
  
where $\hat{\vec{J}}$ represents the angular momentum operator, which in our context corresponds to the neutron's spin ($\hat{\vec{S}}$), thus $j=1/2$.  The states for which we will calculate the expectation values will be defined below, but as the theorem requires, they are angular momentum eigenstates.

It is evident that  $[\hat{\cal{D}}, \hat{\vec{J}}] = 0$, reinforcing the notion that $\hat{\cal{D}}$ is a scalar operator, as implied by the notation. However, it's worth noting that the equality \eqref{effective-id} in terms of operators cannot be regarded as an absolute identity. This is because the components of $\hat{\vec{D}}$ exhibit commutativity among themselves, unlike those of $\hat{\vec{J}}$, which do not follow the same pattern. The significance of this theorem lies in its utility for calculating the essential matrix element terms $\langle\alpha |\hat{\vec{D}}|\beta\rangle$ as required for our computation. 

The following analysis aims to exhibit how the quantum uncertainty of the EDM  enters into the analysis of the weak measurement to ascertain its influence on the accuracy with which the quantity of interest can be measured.

In order to do this, we proceed to write the quantum state of the neutron, explicitly separating the degrees of freedom as indicated above. That is, we express the state of a neutron essentially at rest in the lab and with a spin orientation $\vec{\sigma}$ as:
$ |\psi\rangle =   N \int d(d)   e^{-  (d- d_n)^2/ 2\Delta^2}  |d\rangle \otimes | \vec \sigma\rangle  $, here, $|d\rangle$ represents the eigenstates of $\hat{\cal{D}}$ (normalized according to $\langle d' | d \rangle = \delta(d - d')$), $\tilde{d_n} = \langle n | \hat{\cal{D}} | n \rangle$ signifies the expectation value of the dipole moment scalar in the neutron basis state, and $\Delta$ represents its quantum uncertainty. It's important to note that for any neutron state with a specific orientation, the relevant quantities can be obtained using equation \eqref{effective-id}. The notation $|\vec{\sigma}\rangle$ denotes eigenstates of the spin along the direction $\vec{\sigma}$, where $\vec{\sigma} \cdot \hat{\vec{S}} |\vec{\sigma}\rangle = 1/2 |\vec{\sigma}\rangle$. At this juncture, it is also noteworthy that, due to the Projection Theorem, $\tilde{d_n} = d_n$ \footnote{Here, we have defined $|\expval{\Vec{D}}| \equiv |\Vec{d_n}| = d_n$.}.

Our  analysis  starts  by  writing the  initial  state   representing a pre-selected neutron  (prepared in the lab to  be  essentially  at rest and with  its spin pointing in the direction  $ +z$, namely  
$ |\psi_0\rangle =   N \int d(d)   e^{-  (d- d_n)^2/ 2\Delta^2}  |d\rangle \otimes | z + \rangle  $).   Proper  state normalization implies  $N^2 \int d(d)   e^{-  (d- d_n)^2/ \Delta^2}   =1 $.
            
The neutron is then subjected to the effect of an external electric field pointing in the direction  $ +y $  that is applied during a finite time interval from  $t=0$ to $ t=t_f$,   which we take to be switched on and off slowly enough to avoid transient effects.
          
The interaction of the  neutron with  this  electric field   is  represented by the  Interaction Hamiltonian:
           
\begin{equation}
        H_i =  \vec E (t)\cdot\hat{\vec D}  
\end{equation}   
where, as noted, the electric field is taken to point in the direction $+y$.

We are interested in computing  the probability amplitude  for finding the  neutron in  the   post-selected  state  
$|\psi'\rangle =   N \int d(d)   e^{-  (d- d_n)^2/2 \Delta^2}   |d\rangle \otimes | z - \rangle  $ at $t= t_f$. Thus, we are dealing with a version of weak measurements of the electric dipole moment involving pre and post-selected states \citep{Aharonov}.

The  direct  calculation  of this amplitude is now :
\begin{equation}
\begin{split}
{\cal{ A} } &= \langle \psi'|  {\cal T}  e^{i \int  H_i  dt}  |\psi_0\rangle = \langle \psi'|  {\cal T}  e^{i \int E_y(t) \hat{ D_y } dt}  |\psi_0 \rangle \\
&=\langle \psi'|  e^{i   \hat { D_y }  \int E_y(t)  dt }    N \int d(d)   e^{-  (d- d_n)^2/ \Delta^2}  \\
&\cross |d \rangle \otimes \frac{1}{\sqrt 2} (| y + \rangle + | y -\rangle) 
\label{amplitude}
\end{split}
\end{equation}
where $ {\cal T}  $  stands for the time order product (which will be irrelevant in this treatment where the electric field is taken as classical and with fixed orientation). In the  last line  we have used  $  | z + \rangle  = \frac{1}{\sqrt 2} (| y + \rangle + | y -\rangle) $.
\medskip
     
We  define  now   $ A\equiv  \int_0^{t_f}   E(t)  dt $ and  $ B  \equiv \frac{A}{2 \sqrt{(1/2(1/2+1))}}  $ and   then write

\begin{equation}
\begin{split}
{\cal{ A} } &=   \langle\psi'|  e^{i \int  A  \hat {\cal D}  \hat S_y/ \sqrt{(1/2(1/2+1))} }    N \int d(d)   e^{-  (d- d_n)^2/ \Delta^2}  \\
&\cross |d\rangle \otimes \frac{1}{\sqrt 2} (| y + \rangle + |y-\rangle )\\
&=\langle \psi'|  N \int d(d)    e^{-  (d- d_n)^2/ \Delta^2} \\
&\cross|d\rangle \otimes \frac{1}{\sqrt 2} ( e^{idB}| y + \rangle +e^{-idB} |y-\rangle)
\end{split}
\label{amplitude2}
\end{equation}

Now  we  write
$ \langle \psi'| =   N \int d(d')   e^{-  (d'- d_n)^2/ 2\Delta^2}   \langle d' | \otimes   \frac{1}{\sqrt{2} }(\langle y + | - \langle y-| )  $   so that

\begin{equation}
\begin{split}
 {\cal{ A} }  
 &=  N^2 \int d(d)    e^{-  (d- d_n)^2/ \Delta^2}   \frac{1}{ 2} ( e^{idB} -e^{-idB} )\\
 &=  \frac{1}{ 2} ( e^{id_nB} -e^{-id_nB} ) e^{-B^2 \Delta^2/2} =-i \sin{( d_n B)}e^{-B^2 \Delta^2/2}.
 \label{amplitude-final}
 \end{split}
 \end{equation}

This expression illustrates the feasibility of measuring  $ d_n$,  the expectation value of the Electric Dipole Moment of the neutron, to arbitrarily large precision regardless of the magnitude of the uncertainty $\Delta$. This task is accomplished by measuring the probability of this transition with arbitrarily high accuracy, which can be achieved simply by repeating the experiment sufficiently many times (or using a sufficiently large number of neutrons). 
Note that,  as expected,  if the applied electric field and the neutron electric dipole moment satisfy $ d_n B = n \pi$, the amplitude would vanish as it would correspond to a precession that returned the neutron's spin orientation to its original one. 

A valid concern may arise regarding the expression for the amplitude and, consequently, the probability, as it appears to rely on the variable $\Delta$, which remains, in principle, an unknown quantity. However, the key insight lies in that the term $B$ relies on the applied electric field's magnitude and the duration of its application. Notably, the functional relationship governing this dependence is distinctly determined by the two parameters, $d_n$ and $\Delta$. This distinct functional form enables the separate extraction of values for these two quantities using the dependence on the electric field's characteristics and duration.

In fact, in the limit  where $ B$ is  very small (so that the lowest order  term in  a power expansion can  be taken as reliable),   we have:
 
\begin{equation}
\label{amplitude-final2}
 {\cal{ A} } \approx   -id_nB  e^{-B^2 \Delta^2/2}  
\end{equation}

so that the probability 

\begin{equation}
\label{amplitude-final3}
 {\cal{ P} } \approx    |d_nB|^2  e^{-B^2 \Delta^2} \approx |d_nB|^2  
\end{equation}

can be converted directly into a determination of $d_n$ (assuming $B$, which depends only on the applied electric field,  is known). In practice, the experiments performed up to this date have served to set bounds in such probability, which, in turn,  become bounds on the quantity $d_n$.   We note that this quantity is positive definite irrespective of the sign of $ d_n$  (of whether the  EDM is aligned in the same direction as the spin or the opposite one). 
It is worth mentioning that under the condition where   $ B $ is not small enough,  the effect of the electric dipole moment's quantum uncertainty   $\Delta $, as shown in \eqref{amplitude-final}  above,  reduces the probability by a factor $e^{-B^2 \Delta^2} $.  This is not a very intuitive feature, and a deeper understanding would probably require a more general analysis, which is out of the scope of the present manuscript.

\section{Discussion and conclusions}

The recognition that there is a profound difference between a strong measurement of a physical quantity—a process that often involves altering the state of the system as a result of the measuring process—and a weak measurement of the expectation value of that quantity, making use of one or several auxiliary systems (\textit{ancillas}) that are made to interact very weakly with the system of interest and which lead to negligible changes in the state of the system, represents, we believe, the solution to what seemed as a serious discrepancy between the magnitude of the quantum uncertainty and the bounds that are extracted from the experiments on the nEDM.

We started with something that seemed to be a conflict between experimental data and what quantum theory itself was expected to allow to be measured and ended with what we think is a clear understanding of what is going on.
 
Moreover, we think several lessons can be taken to heart from this analysis; some of these are well-known general lessons that are sometimes overlooked in practice. First, we should not confuse the value of a physical observable  $ O$ represented by the operator $  \hat  O$ and its expectation value  $  \langle \psi | \hat  O | \psi  \rangle$ in a certain state $ | \psi  \rangle$.    According to quantum theory, an observable only has a definite value for an observable  $ O$ if its state is one of the eigenstates of that operator corresponding to one of the eigenvalues  $\lbrace o_i \rbrace$  of the observable,  that is, if $| \psi  \rangle$   is such that   $\hat  O | \psi  \rangle = o_i | \psi  \rangle$.  The expectation value, on the other hand,  is always a well-defined quantity $  \langle \psi | \hat  O | \psi  \rangle$, and there does not seem to be any basic principle preventing its measurement with arbitrarily high precision. In fact one can do that by preparing a large enough number of copies of the system,    preparing them all in the same state,   and performing on each one a strong measurement.  Such strong measurements, in general, do alter seriously the individual system's state. The point, however, is that in such a scheme, one expects the quantum uncertainty of the initial state to show up as a dispersion in the distribution of results that should translate into statistical errors emerging from the experiment analysis.
This issue could become particularly problematic when, instead of determining a finite value, one is attempting to place bounds on a quantity that is compatible with zero.  In that case,  the statistical dispersion can be expected to overwhelm the sought-for bounds.  That is the basis of the puzzling situation we were describing at the beginning of this work. However, the quantity $  \langle \psi | \hat  O | \psi  \rangle$  can also be measured by other means, as illustrated in the weak measurement process discussed in this work.  In the case at hand,  and although people often talk about measuring the nEDM,  motivated by the requirement for it vanish assuming  CP symmetry of the strong  interactions\footnote{ As everywhere else in our discussion, we are ignoring the effects of   CP  violation in the electro-weak sector.},  what we have,   in fact,  are measurements of the expectation value of the nEDM.  Moreover, it should be emphasized that the quantity that is required to vanish by CP symmetry of the strong interactions is precisely such expectation value.  The CP  symmetry does not require the neutron to be an actual eigenstate with the vanishing eigenvalue of the    EDM  operator.

In this sense, it is worth considering the possibility that novel and more refined experiments might eventually establish a clear non-vanishing value of  $ d_n $,  which might arise from either the standard strong CP-violating parameter $\theta$ or from other physics beyond the standard model -- for instance  Supersymmetric  models\citep{Pospelov_2005, RAMSEYMUSOLF_2008},   left-right symmetric models \citep{PhysRevD.10.275, PhysRevD.11.703.2, PhysRevD.11.566, PhysRevD.12.1502}, or the two-Higgs model \citep{Inoue_2014}--. In that eventuality, we might be hard-pressed to ascertain the exact origin of the phenomena. In this context, it is interesting to note that, as indicated by our analysis,    specifically by equations \eqref{amplitude-final} and \eqref{amplitude-final2}, the relevant amplitude for hard photon emission depends both on the parameters   $ d_n$, the expectation value of the electric dipole moment of the neutron but also on the uncertainty  $\delta$.  Moreover,   such dependence has a form that allows these two quantities to be disentangled by suitable variation of the experimentally controllable quantity $ A = \int  E  dt $. It, thus,  would be interesting to explore what the various models leading to a non-vanishing  $ d_n$ predict regarding the value of $ \Delta$, as the two quantities might be extracted from suitably adapted experiments.

We hope that our discussion clarifies what seemed, at first glance, to be a rather puzzling situation.

\section*{Acknowledgements}
We acknowledge helpful discussions with Prof.  David  Albert, Prof.  Lev Vaidman,   and Prof. Alejandro Perez. D.S.  received partial support from the Conahcyt grant  140630,  and  PAPIIT- UNAM grant  IG 100124-. L.B.P acknowledges the support of PAPIIT-UNAM grant AG102023.

\bibliographystyle{elsarticle-harv} 

\bibliography{biblio}

\end{document}